\documentclass{revtex4}
\usepackage{epsfig}
\usepackage{amssymb}
\usepackage{amsmath}
\usepackage{amsfonts}
\usepackage{graphicx}
\usepackage{mathrsfs}
\usepackage[dvips]{color}
\usepackage{multirow}

\newcommand{\R}{\mathbb{R}}
\newcommand{\C}{\mathbb{C}}

\newcommand{\fc}{\mathfrak{c}}

\newcommand{\fn}{\mathfrak{n}}

\newcommand{\fz}{\mathfrak{z}}

\newcommand{\cO}{\mathcal{O}}

\newcommand{\be}{\begin{equation}}
\newcommand{\ee}{\end{equation}}
\newcommand{\bea}{\begin{eqnarray}}
\newcommand{\eea}{\end{eqnarray}}
\newcommand{\nn}{\nonumber}

\newcommand{\ed}{\end{document}}

\newcommand{\bi}{\begin{itemize}}
\newcommand{\ei}{\end{itemize}}

\newcommand{\bce}{\begin{center}}
\newcommand{\ece}{\end{center}}

\newcommand{\IM}{\,{\rm Im}}

\newcommand{\mf}{\mathfrak{n}_0}
\newcommand{\mg}{\mathfrak{g}_0}

\begin{document}
\title{Perturbative Analysis of Spectral Singularities and Their Optical Realizations}
\author{Ali~Mostafazadeh and Saber Rostamzadeh}
\address{Departments of Mathematics and Physics, Ko\c{c}
University,\\ Sar{\i}yer 34450, Istanbul, Turkey\\
amostafazadeh@ku.edu.tr}

\begin{abstract}

We develop a perturbative method of computing spectral singularities of a Schr\"odinger operator defined by a general complex potential that
vanishes outside a closed interval. These can be realized as zero-width resonances in optical gain media and correspond to a lasing effect
that occurs at the threshold gain. Their time-reversed copies yield coherent perfect absorption of light that is also known as antilasering. We use our general results to establish the exactness of the $n$-th order perturbation theory for an arbitrary complex potential consisting of $n$ delta-functions, obtain an exact expression for the transfer matrix of these potentials, and examine spectral singularities of complex barrier potentials of arbitrary shape. In the context of optical spectral singularities, these correspond to inhomogeneous gain media.
\medskip

\hspace{6.2cm}{Pacs numbers: 03.65.-w, 03.65.Nk, 42.25.Bs,
24.30.Gd}
\end{abstract}

\maketitle

\section{Introduction}

A spectral singularity is a well-known mathematical concept \cite{ss-math} with an interesting physical counterpart; it corresponds to a
zero-width resonance \cite{prl-2009}. As shown in Refs.~\cite{prl-2009,pra-2009}, such resonances can be realized in optical systems consisting of a gain medium. They give rise to a particular lasing effect that occurs at the threshold gain \cite{pra-2011a}. Since
the publication of \cite{prl-2009} there has been a growing interest in the study of the physical applications of spectral singularities
\cite{pra-2009,pra-2011a,ss-phys,longhi-phys-10,pra-2011b,pla-2011}. In particular, it turns out that a time-reversed copy of an optical spectral singularity (OSS) \cite{prl-2009} corresponds to a coherent perfect absorption of light  \cite{longhi-phys-10}. This is the basic physical phenomenon occurring in an antilaser \cite{antilaser}.

The investigation of OSSs that is conducted in \cite{prl-2009,pra-2009,pra-2011a} relies on the assumption that the gain (attenuation) coefficient is constant throughout the gain (loss) region(s). This is a simplifying condition that is almost impossible to fulfill in practice. In order to have more realistic models displaying an OSS we need to develop computational techniques for the cases that the optically active medium is inhomogeneous. The first step in this direction is taken in \cite{pra-2011b}, where a semiclassical method of calculating spectral singularities is employed. The purpose of the present paper is to develop an alternative method of exploring spectral singularities that is based on perturbation theory. This is especially useful, because for most optically active media, the imaginary part of the effective potential, that is responsible for the emergence of an OSS, is several orders of magnitude smaller than its real part \cite{footnote}.

Consider the time-independent Sch\"odinger equation
    \be
    -\psi''(x)+v(x)\psi(x)=k^2\psi(x),
    \label{sch-eq}
    \ee
where $x\in\R$ and $v$ is a complex-valued potential that vanishes outside the interval $[0,1]$. We can express the general solution of (\ref{sch-eq}) as
    \be
    \psi(x)=\left\{\begin{array}{ccc}
    A_- e^{ikx}+B_-e^{-ikx}&{\rm for}&x<0,\\
    A_0\,\phi_1(x;k)+B_0\,\phi_2(x;k)&{\rm for}&0\leq x\leq 1,\\
    A_+ e^{ikx}+B_+e^{-ikx}&{\rm for}&x>1,\end{array}\right.
    \ee
where $A_\pm,B_\pm,A_0,B_0$ are complex coefficients and $\phi_1(\cdot;k)$ and $\phi_2(\cdot,k)$ are solutions of (\ref{sch-eq}) on the interval $[0,1]$ that are determined by the initial conditions \cite{pra-2011b}:
    \be
    \phi_1(0;k)=1,~~~\phi_1'(0;k)=-ik,~~~
    \phi_2(0;k)=1,~~~\phi_2'(0;k)=0.
    \label{ini}
    \ee

Spectral singularities are given by the real zeros of the $M_{22}$ entry of the transfer matrix $\mathbf{M}$ of the system \cite{prl-2009}. This is the $2\times 2$ matrix $\mathbf{M}$ that satisfies $\left[\begin{array}{c} A_+\\B_+\end{array}\right]=
\mathbf{M}\left[\begin{array}{c}A_-\\B_-\end{array}\right]$. Demanding that $\psi$ is a continuously differentiable function (throughout $\R$) and using (\ref{ini}), we find the following expression for the transfer matrix.
    \be
    \mathbf{M}=\frac{1}{2ik}\left[
    \begin{array}{cc}
    -e^{-ik}[\Gamma_{1+}(k)-2\Gamma_{2+}(k)] &~~~~e^{-ik}\Gamma_{1+}(k)\\
    &\\
    e^{ik}[\Gamma_{1-}(k)-2\Gamma_{2-}(k)]& -e^{ik}\Gamma_{1-}(k)\end{array}\right],
    \label{transfer-matrix}
    \ee
where
    \be
    \Gamma_{j\pm}(k):=
    \phi_j'(1;k)\pm ik\phi_j(1;k).
    \label{jost}
    \ee
In view of (\ref{transfer-matrix}) spectral singularities are the real zeros of the Jost function $\Gamma_{1-}$, \cite{pra-2011b}.

In order to develop a perturbative method of computing spectral singularities we consider potentials of the form
    \be
    v(x):=\left\{\begin{array}{cc}
    v_0(x)+\epsilon\, v_1(x)&{\rm for}~0\leq x\leq 1,\\ &\\
    0 & {\rm otherwise},\end{array}\right.
    \label{v=}
    \ee
where $v_0$ is an exactly solvable potential, $\epsilon$ is a real perturbation parameter, and $v_1$ is an arbitrary potential. Our aim is to use perturbation theory to compute the solution $\phi_1(\cdot;k)$ of the Schr\"odinger equation (\ref{sch-eq}), the Jost function $\Gamma_{1-}$, and its real zeros.

The plan of the paper is as follows. In Section~2 we derive a perturbative series expansion for $\phi_1(\cdot;k)$ and discuss how it can be used to locate the spectral singularities. In Section~3 we use our perturbative method to examine  potentials involving one or more delta functions. Here we establish the exactness of the perturbation theory and give an explicit formula for the transfer matrix of the system. In Section~4, we apply our method to an arbitrary piecewise continuous complex barrier potential. In Section~5, we use the results of Section~4 to study the OSS of an infinite planar slab gain medium whose gain/loss coefficient varies along the normal direction. Finally, in Section~6 we give a summary of our findings and concluding remarks.

\section{Perturbative Calculation of Spectral Singularities}

We start our analysis by introducing the differential operator
    \be
    L:=\frac{d^2}{dx^2}-v_0(x)+k^2,
    \label{L=}
    \ee
and note that $\phi_1(\cdot;k)$ and $\phi_2(\cdot;k)$ are solutions of the differential equation
    \be
    L\phi(x)=\epsilon v_1(x) \phi(x),
    \label{3-5}
    \ee
on the interval $[0,1]$ that are uniquely determined by the initial conditions (\ref{ini}). To construct a perturbative solution of (\ref{3-5}) we insert the ansatz
     \begin{equation}
     \phi(x) =\sum_{j=0}^{\infty}\phi^{\scriptscriptstyle{(j)}}(x)\,\epsilon^j
     \label{3-6}
     \end{equation}
in (\ref{3-5}) and demand that it holds term by term in powers of $\epsilon$. This yields
    \bea
    L\phi^{(0)}(x)&=&0,
    \label{zeroth-order}\\
    L\phi^{\scriptscriptstyle{(\ell)}}(x)&=&
    v_1(x)\phi^{\scriptscriptstyle{(\ell-1)}}(x),
    \label{3-7}
    \eea
where $\ell=1,2,3,\cdots$. We can express the general solution of the latter equation in the form
    \begin{equation}
    \phi^{\scriptscriptstyle{(\ell)}}(x)=a_\ell\, \phi^{\scriptscriptstyle{(0)}}_1(x)+
    b_\ell\,\phi^{\scriptscriptstyle{(0)}}_2(x)+
    \int_0^x G(x,y)v_1(y)
    \phi^{\scriptscriptstyle{(\ell-1)}}(y)dy,
    \label{G1}
    \end{equation}
where $a_\ell,b_\ell$ are constant coefficients, $\phi^{(0)}_1$ and $\phi^{(0)}_2$ are linearly-independent solutions of (\ref{zeroth-order}), and $G$ is the Green's function for the operator $L$ that can be expressed as \cite{Diprima}:
    \be
    G(x,y) =\frac{\phi^{(0)}_1(y)\phi^{(0)}_2(x)-
    \phi^{(0)}_2(y)\phi^{(0)}_1(x)}{\phi^{(0)}_1(y){\phi^{(0)}_2}'(y)-
    \phi^{(0)}_2(y){\phi_1^{{(0)}}}'(y)}.
    \ee

A convenient choice, that determines $\phi^{(0)}_1$ and $\phi^{(0)}_2$ in a unique manner, is to demand that they also satisfy the initial conditions (\ref{ini}), i.e.,
    \be
    \phi^{(0)}_1(0)=1,~~~~{\phi^{(0)}_1}'(0)=-ik,~~~~
    \phi^{(0)}_2(0)=1,~~~~{\phi^{(0)}_2}'(0)=0.
    \label{ini2}
    \ee
In this case the denominator of the Green's function that coincides with the Wronskian of $\phi^{(0)}_1$ and $\phi^{(0)}_2$ takes the value $ik$, and we find
    \be
    G(x,y) =ik^{-1}\left[\phi^{(0)}_1(x)\phi^{(0)}_2(y)-
    \phi^{(0)}_2(x)\phi^{(0)}_1(y)\right].
    \label{green}
    \ee
Another advantage of imposing (\ref{ini2}) is that it identifies the zeroth-order term in the perturbative expansion of $\phi_j(\cdot;k)$ with $\phi^{(0)}_j(\cdot;k)$. Moreover, if we denote the higher order terms in this expansion by $\phi^{(\ell)}_j(\cdot;k)\,\epsilon^\ell$ (with $\ell\geq 1$), so that
    \be
    \phi_j(x;k)=\sum_{\ell=0}^\infty \phi^{(\ell)}_j(x;k)\,\epsilon^\ell,~~~~~
     \phi^{(0)}_j(x;k)= \phi^{(0)}_j(x),
    \label{3-21}
    \ee
then (\ref{ini}) and (\ref{ini2}) imply that $\phi^{(\ell)}_j(0;k)={\phi^{(\ell)\,\prime}_j}(0;k)=0$ for $\ell\geq 1$.
A direct consequence of this relation is that the coefficients $a_\ell$ and $b_\ell$ of (\ref{G1}) vanish, and we can use (\ref{G1}) to derive the formula:
    \be
    \phi^{(\ell)}_j(x_\ell;k)=\int_0^{x_\ell} dx_{\ell-1}\int_0^{x_{\ell-1}}dx_{\ell-2}\cdots\int_0^{x_1} dx_0
    \:\phi^{(0)}_j(x_0)\prod_{m=1}^\ell G(x_m,x_{m-1})v_1(x_{m-1}),
    \label{3-22}
    \ee
that holds for all $\ell\geq 1$ and $x_\ell\in[0,1]$.

We can use (\ref{3-21}) to obtain a perturbative expansion for the Jost functions (\ref{jost}):
    \be
    \Gamma_{j\pm}(k)=\sum_{\ell=0}^\infty
    \Gamma_{j\pm}^{(\ell)}(k)\,\epsilon^\ell.
    \label{3-23}
    \ee
In view of (\ref{jost}), (\ref{3-21}), and (\ref{3-22}), we have
    \bea
    \Gamma_{j\pm}^{(0)}(k)&=&{\phi_j^{(0)}}'(1)\pm ik \phi^{(0)}_j(1),
    \label{3-24}\\
    \Gamma_{j\pm}^{(\ell)}(k)&=&\int_0^1 dx_{\ell}\int_0^{x_{\ell}}dx_{\ell-1}\cdots\int_0^{x_2} dx_1
    \:\phi^{(0)}_j(x_1)\times\nn\\
    &&
    \Big\{\left[G'(1,x_{\ell})\pm ik G(1,x_{\ell})\right]v_1(x_{\ell})
    \prod_{m=1}^{\ell-1} G(x_{m+1},x_{m})v_1(x_{m})\Big\},
    \label{3-25}
    \eea
where
    \be
    G'(x,y):=\partial_x G(x,y)=ik^{-1}\left[{\phi_1^{(0)}}'(x)\phi^{(0)}_2(y)-
    {\phi_2^{(0)}}'(x)\phi^{(0)}_1(y)\right].
    \label{green2}
    \ee

Next, we recall that in general $v_0$ and $v_1$ involve a set of complex coupling constants $(\fz_1,\fz_2,\cdots,\fz_n)$. These together with the perturbation parameter $\epsilon$ are the parameters that enter in the expression (\ref{3-23}) for the Jost solutions. Spectral singularities are given by the real values of $k$ for which
    \be
    \Gamma_{1-}(k)=0.
     \label{eqn}
     \ee
This is a complex equation involving $n$ complex variables, $\fz_1,\fz_2,\cdots,\fz_n$, and two real variables, $\epsilon,k$.

If the unperturbed potential $v_0$ is real-valued or more generally does not support a spectral singularity, the emergence of spectral singularities is a consequence of the perturbation $\epsilon\, v_1$. Otherwise the presence of the latter leads to small changes in the location of the spectral singularities of $v_0$ that are given by the real solutions of
    \be
    \Gamma_{1-}^{(0)}(k)=0.
     \label{eqn-0}
     \ee
In this case, we can again adopt perturbative theory to construct solutions of (\ref{eqn}) using those of (\ref{eqn-0}),  \cite{endnote-x1}. Before, we explore the details of this construction, we consider a situation where perturbation theory yields the exact solution of the problem.

\section{Array of Complex Delta-Function Potentials}

Consider the case that $v_0(x)=0$ and the perturbation involves an array of Dirac delta-functions \cite{dirac-delta,Griffiths, jpa-2006,jpa-2009}:
    \be
    v_1(x)=\sum_{i=1}^n \fz_i\delta(x-a_i),
    \label{array}
    \ee
where $\fz_1,\fz_2,\cdots,\fz_n$ are complex coupling constants, and $a_1,a_2,\cdots,a_n$ are arbitrary positive numbers satisfying
    \be
    0<a_1<a_2<\cdots<a_n<1.
    \label{order}
    \ee

Because $v_0=0$, Eqs.~(\ref{L=}), (\ref{zeroth-order}), (\ref{ini2}), (\ref{green}), and (\ref{green2}) give
    \bea
    &&\phi_1^{(0)}(x)=e^{-ikx},~~~~~~~~~~~~~~~~~~~\phi_2^{(0)}(x)=\cos(kx),
    \label{4-1}\\
    &&G(x,y)=\frac{\sin[k(x-y)]}{k},~~~~~~~~G'(x,y)=\cos[k(x-y)].
    \label{4-2}
    \eea
If we substitute (\ref{array}), (\ref{4-1}), and (\ref{4-2}) in (\ref{3-22}), and use the properties of the delta-function to perform the relevant integrals, we find
    \be
    \phi^{(\ell)}_j(x;k)=\sum_{i=1}^n Z^{(\ell)}_{ji}\sin[k(x-a_i)]\theta(x-a_i),
    \ee
where $\ell\geq 1$, for all $i_{\ell}=1,2,\cdots,n$ and $j=1,2$,
    \bea
    Z_{ji_\ell}^{(\ell)}:=k^{-\ell}\fz_{i_\ell}\!\!\!\!
    \sum_{i_1,i_2,\cdots,i_{\ell-1}=1}^n\!\!\!\!\phi_j^{(0)}(a_{i_1})\,
    \prod_{m=1}^{\ell-1} \fz_{i_{m}}
    \sin[k(a_{i_{m+1}}-a_{i_{m}})]\theta(a_{i_{m+1}}-a_{i_{m}}),
    \label{W}
    \eea
and $\theta$ stands for the Heaviside step function,
    \be
    \theta(x)=\left\{\begin{array}{ccc}
    1 &{\rm for}& x\geq0,\\
    0 &{\rm for}& x<0.\end{array}\right.
    \label{theta}
    \ee
In view of (\ref{order}) and (\ref{theta}), the product on the right-hand side of (\ref{W}) vanishes identically, if $i_{m+1}\leq i_{m}$. Hence
    \be
    Z_{ji_\ell}^{(\ell)}(k)=\theta(n-\ell)\:
    k^{-\ell}\fz_{i_\ell}\!\!\!\!
    \sum_{i_1<i_2<\cdots<i_{\ell-1}=1}^{i_\ell}\!\!\!\!
    \phi_j^{(0)}(a_{i_1})\,
    \prod_{m=1}^{\ell-1} \fz_{i_{m}}
    \sin[k(a_{i_{m+1}}-a_{i_{m}})],
    \label{Z2}
    \ee
and as a result $\phi_j^{(\ell)}(x;k)=0$ for all $\ell>n$.
This proves the following theorem.
    \begin{itemize}
    \item[]\textbf{Theorem}: \emph{For a point interaction consisting of $n$ delta-functions with arbitrary centers and possibly complex coupling constants, the $n$-th order perturbation theory is exact.}
    \end{itemize}

Next, we compute the Jost functions (\ref{3-23}). Using (\ref{array}), (\ref{4-1}), and (\ref{4-2}) in (\ref{3-24}) and (\ref{3-25}), we find
    \bea
    \Gamma_{1+}^{(0)}(k)&=&0,~~~~~~~~\Gamma_{1-}^{(0)}(k)=-2i e^{-ik},
    ~~~~~~~~\Gamma_{2\pm}^{(0)}(k)=\pm i k e^{\pm ik},
    \label{4-31}\\
    \Gamma_{j\pm}^{(\ell)}(k)&=& \theta(n-\ell)\:
    \left(\frac{\pm 1}{2 i k}\right)^{\ell-1}\!\!
    \sum_{i_1<i_2<\cdots<i_{\ell}=1}^{n}\!\!\!\!
    \Omega_{i_1 j \pm } \prod_{p=1}^\ell\fz_{i_p}
    \prod_{m=1}^{\ell-1}
    \left[1-e^{\mp 2ik(a_{i_{m+1}}-a_{i_m})}\right],
    \label{4-32}
    \eea
where $\ell\geq 1$, and for all $i_1=1,2,\cdots,\ell$,
    \be
    \Omega_{i_1 1 +}:=e^{ik(1-2a_{i_1})},~~~~~~~~~
    \Omega_{i_1 1 -}:=e^{-ik},~~~~~~~~~
    \Omega_{i_1 2 \pm}:=\frac{1}{2}\left(1+e^{\mp2ika_{i_1}}\right)e^{\pm ik}.
    \ee
Substituting (\ref{4-1}) in this relation, using the result in (\ref{transfer-matrix}), and setting $\epsilon=1$, we obtain the following expressions for the entries of the transfer matrix corresponding to the potential (\ref{array}):
    \bea
    M_{11}&=&1+\sum_{\ell=1}^n(2ik)^{-\ell}
    \!\!\!\!\!\! \sum_{i_1<i_2<\cdots<i_{\ell}=1}^{n}
    \prod_{p=1}^\ell\fz_{i_p}\prod_{m=1}^{\ell-1}
    \left[1-e^{- 2ik(a_{i_{m+1}}-a_{i_m})}\right]
    \label{M11}\\
    M_{12}&=&\sum_{\ell=1}^n(2ik)^{-\ell}
    \!\!\!\!\!\! \sum_{i_1<i_2<\cdots<i_{\ell}=1}^{n}e^{-2ika_{i_1}}
    \prod_{p=1}^\ell\fz_{i_p}\prod_{m=1}^{\ell-1}
    \left[1-e^{- 2ik(a_{i_{m+1}}-a_{i_m})}\right]
    \label{M12}\\
    M_{21}&=&\sum_{\ell=1}^n (-2ik)^{-\ell}
    \!\!\!\!\!\! \sum_{i_1<i_2<\cdots<i_{\ell}=1}^{n}e^{2ika_{i_1}}
    \prod_{p=1}^\ell\fz_{i_p}\prod_{m=1}^{\ell-1}
    \left[1-e^{2ik(a_{i_{m+1}}-a_{i_m})}\right]
    \label{M21}\\
    M_{22}&=&1+\sum_{\ell=1}^n(-2ik)^{-\ell}
    \!\!\!\!\!\! \sum_{i_1<i_2<\cdots<i_{\ell}=1}^{n}
    \prod_{p=1}^\ell\fz_{i_p}\prod_{m=1}^{\ell-1}
    \left[1-e^{2ik(a_{i_{m+1}}-a_{i_m})}\right].
    \label{M22}
    \eea

It is instructive to examine the cases $n=1,2$.

For $n=1$, we have $v_1=\fz_1\delta(x-a_1)$, and (\ref{M11}) -- (\ref{M22}) give
    \be
    M_{11}=1-\frac{i\fz_1}{2k},~~~~~~
    M_{12}=-\frac{i\fz_1\,e^{-2ika_1}}{2k},~~~~~~
    M_{21}=\frac{i\fz_1\,e^{2ika_1}}{2k},~~~~~~M_{22}=1+\frac{i\fz_1}{2k}.
    \ee
This agrees with the results of \cite{pjp-2009}. In particular, a spectral singularity occurs for imaginary values of $\fz_1$ and is given by $k=-i\fz_1/2$, as envisaged in \cite{jpa-2006} and shown in \cite{jpa-2009}.

For $n=2$, we have $v_1=\fz_1\delta(x-a_1)+\fz_2\delta(x-a_2)$, and (\ref{M11}) -- (\ref{M22}) give
    \bea
    M_{11}&=&1-\frac{i(\fz_1+\fz_2)}{2k}-\frac{\fz_1\fz_2\left[1- e^{-2ik(a_2-a_1)}\right]}{4k^2},\\
    M_{12}&=&-\frac{i(\fz_1 e^{-2ik a_1} +\fz_2e^{-2ik a_2})}{2k}
    -\frac{\fz_1\fz_2(e^{-2ika_1}-e^{-2ika_2})}{4k^2},\\
    M_{21}&=&\frac{i(\fz_1 e^{2ik a_1} +\fz_2e^{2ik a_2})}{2k}
    -\frac{\fz_1\fz_2(e^{2ika_1}-e^{2ika_2})}{4k^2},\\
    M_{22}&=&1+\frac{i(\fz_1+\fz_2)}{2k}-
    \frac{\fz_1\fz_2\left[1-e^{2ik(a_2-a_1)}\right]}{4k^2}.
    \label{double-delta}
    \eea
Spectral singularities are therefore given by the real values of $k$ for which the right-hand side of (\ref{double-delta}) vanishes. Again this is in complete agreement with the results of \cite{jpa-2009}.

For the cases that $a_i:=i/(n+1)$ and $\fz_1=\fz_2=\cdots=\fz_n$, the potential $v$ is locally periodic, and we can use the results of \cite{Griffiths} to compute the transfer matrix of the system. We have checked by explicit calculation for small values of $n$ that (\ref{M11}) -- (\ref{M22}) give the same expression for the transfer matrix as the one derived in \cite{Griffiths}.

\section{Complex Barrier Potentials}

Consider the potentials of the form (\ref{v=}) where for all $x\in[0,1]$,
    \be
    v_0(x):=\fz_1, ~~~~~~v_1(x):=\fz_2 f(x),
    \label{5-1}
    \ee
$\fz_1$ and $\fz_2$ are complex coupling constants, $f:[0,1]\to\C$ is an integrable function satisfying $\int_0^1dx|f(x)|\leq 1$, and  $|\epsilon\fz_2|\ll |\fz_1|$. As explained in Ref.~\cite{pra-2011b}, these potentials appear in the study of the OSS of an infinite planar slab gain medium with gain coefficient changing along the normal direction to the slab.

In order to determine the spectral singularities of the potentials given by (\ref{v=}) and (\ref{5-1}), we first use (\ref{L=}), (\ref{zeroth-order}), (\ref{ini2}), (\ref{green}), and (\ref{green2}) to compute
    \bea
    &&\phi_1^{(0)}(x)=\cos(\fn k x)-i\fn^{-1}\sin(\fn k x),~~~~
    \phi_2^{(0)}(x)=\cos(\fn k x),
    \label{5-2}\\
    &&G(x,y)=(\fn k)^{-1}\sin[\fn k(x-y)],~~~~
    G'(x,y)-ik G(x,y)=\phi_1^{(0)}(x-y),
    \label{5-3}
    \eea
where
    \be
    \fn:=\sqrt{1-\frac{\fz_1}{k^2}}.
    \label{fn=}
    \ee
Clearly, $\fn=0$ marks a singularity of our construction that we will avoid.
If we substitute (\ref{5-2}) in (\ref{3-24}), we find
    \bea
    \Gamma_{1-}^{(0)}(k)=F_0(\fn,k)&:=&
    -\fn^{-1}k\left[(\fn^2+1)\sin(\fn k)+2i\fn\cos(\fn k)\right]\nn\\
    &=&\frac{k (\fn+1)^2 e^{i\fn k}}{2i\fn}\left[ e^{-2i\fn k}-
    \left(\frac{\fn-1}{\fn+1}\right)^2\right].
    \label{5-4}
    \eea
This is consistent with the results of Ref.~\cite{pra-2011a}, because it implies that for $\epsilon=0$, that corresponds to a constant complex barrier potential, the spectral singularities are determined by the equation:
    \be
    e^{-2i\fn k}-\left(\frac{\fn-1}{\fn+1}\right)^2=0.
    \label{ss-zero}
    \ee
Similarly, using (\ref{3-25}), (\ref{5-2}) and (\ref{5-3}), we obtain for all $\ell\geq 1$,
    \be
    \Gamma_{1-}^{(\ell)}(k)=\fz_2^\ell F_\ell(\fn,k)
    \label{5-5}
    \ee
where
    \bea
    F_\ell(\fn,k)&:=&(\fn k)^{1-\ell}
    \int_0^1\!\!\! dx_\ell\int_0^{x_\ell} \!\!\! dx_{\ell-1}\cdots\int_0^{x_2} \!\!\! dx_1
    \xi(\fn,k,x_1)\prod_{m=1}^{\ell-1}\sin[\fn k(x_{m+1}-x_m)]\prod_{p=1}^\ell f(x_p),~~~~~~
    \label{5-5b}\\
    \xi(\fn,k,x)&:=&\phi_1^{(0)} (x) \phi_1^{(0)}(1-x)\nn\\
    &=&\frac{1}{2}\left\{
    \left(1+\frac{1}{\fn^2}\right)\cos(\fn k)-\frac{2i\sin(\fn k)}{\fn}+\left(1-\frac{1}{\fn^2}\right)\cos[\fn k(2x-1)]\right\}.
    \label{xi=}
    \eea
In particular,
    \bea
    F_1(\fn,k)=\int_0^1 dx\, \xi(\fn,k,x)f(x).
    \eea

In the remainder of this section we explore the application of perturbation theory for treating the following equation whose real solutions yield the spectral singularities.
    \be
    \Gamma_{1-}(k)=\sum_{\ell=0}^\infty F_\ell(\fn,k)\,\fz_2^\ell\,\epsilon^\ell=0.
    \label{5-7}
    \ee

Suppose that we have a spectral singularity for the unperturbed potential, i.e., when $\fz_2=0$. Let $(\fn_0,k_0)$ be the value of $(\fn,k)$ at which this spectral singularity is realized. If we turn on the perturbation,
i.e., set $\epsilon\fz_2\neq 0$, this spectral singularity occurs for a new value $(\fn_\star,k_\star)$ of $(\fn,k)$. Our aim is to express $k_\star$ and $\fn_\star$ as power series in the perturbation parameter $\epsilon$,
    \be
    k_\star=\sum_{m=0}^\infty k_m\epsilon^m,~~~~~~~~~
    \fn_\star=\sum_{m=0}^\infty \fn_m\epsilon^m,
    \label{5-19}
    \ee
and determine the coefficients $k_m$ and $\fn_m$, that respectively take real and complex values. To do this we expand $F_\ell(\fn,k)$ in a power series about $(\fn_0,k_0)$,
    \bea
    &&F_\ell(\fn,k)=\sum_{p,q=0}^n F_{\ell p q}(\fn-\fn_0)^p(k-k_0)^q,
    \label{5-20}\\
    &&F_{\ell p q}:=\frac{1}{p!q!}\frac{\partial^{p+q}
    F_{\ell}(\fn_0,k_0)}{\partial\fn_0^p\partial k_0^q},
    \label{5-21}
    \eea
substitute $k=k_\star$ and $\fn=\fn_\star$ in (\ref{5-7}), and use (\ref{5-19}) and (\ref{5-20}) to express the resulting equation in the form
    \be
    \sum_{j=1}^\infty \fc_j\epsilon^j=0,
    \label{5-22}
    \ee
where $\fc_j$ are complex coefficients depending on $k_m$, $\fn_m$, and $\fz_2$. For example,
    \bea
    \fc_1&=&F_{0 1 0}\,\fn_1+F_{0 0 1}\, k_1+F_{1 0 0} \,\fz_2,
    \label{c1=}\\
    \fc_2&=&F_{0 1 0}\,\fn_2+F_{0 0 1}\, k_2+
    F_{0 2 0}\,\fn_1^2+ F_{0 1 1}\,\fn_1 k_1 +F_{002}\, k_1^2 \nn\\
    &&+(F_{1 1 0}\,\fn_1+F_{1 0 1}\,k_1)\fz_2+F_{200}\,\fz_2^2.
    \label{c2=}
    \eea
Finally, we enforce (\ref{5-22}) by demanding that $\fc_j=0$ for all $j\geq 1$. Because $\fc_j$ involves $k_m$ and $\fn_m$ with $m\leq j$, in this way we obtain an infinite set of algebraic equations for $k_m$ and $\fn_m$ that we can solve iteratively.

For example, in view of (\ref{c1=}) and (\ref{c2=}), $\fc_1=0$ and $\fc_2=0$ give
    \bea
    F_{0 1 0}\,\fn_1+F_{0 0 1}\, k_1&=&-F_{1,0,0}\,\fz_2,
    \label{5-31}\\
    F_{0 1 0}\,\fn_2+F_{0 0 1}\, k_2&=&- \left(
    F_{0 2 0}\,\fn_1^2+ F_{0 1 1}\,\fn_1 k_1 +F_{002}\, k_1^2\right)\nn\\
    &&-(F_{1 1 0}\,\fn_1-F_{1 0 1}\,k_1)\fz_2-F_{200}\,\fz_2^2,
    \label{5-32}
    \eea
respectively.  In general if $\fn$ and $k$ are independent complex and real variables, we can choose $k_m=0$ and use the above method to compute $\fn_m$. However, as we see in the next section, there are situations of physical interest where $\fn$ depends on $k$ and another real variable $g$. In this case, our method produces a perturbative calculation of $k$ and $g$.

\section{OSSs of an Inhomogeneous Planar Slab Gain Medium}

\begin{figure}
    \begin{center}
    \includegraphics[scale=.6,clip]{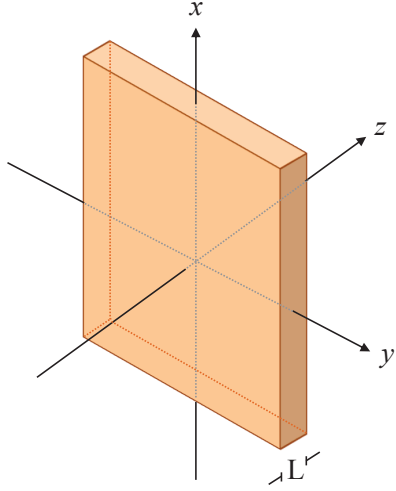}
    \caption{(Color online) Schematic view of an infinite planar slab of gain material of thickness $L$ that is aligned in the $x$-$y$ plane.\label{fig1}}
    \end{center}
    \end{figure}

A simple model supporting OSSs is an infinite planar slab gain medium \cite{pra-2009,pra-2011a}. Suppose we choose a coordinate system in which the slab is aligned parallel to the $x$-$y$ plane and has a thickness $L$, as demonstrated in Figure~\ref{fig1}.
Then the wave equation associated with this system admits a solution that propagates along the $z$-axis and corresponds to the following expression for the electric field \cite{pra-2011b}.
    \[\vec E(z,t)=E_0 e^{-i\omega t}\Psi(z)\hat e_x,\]
where $E_0$ is a constant coefficient, $\omega$ is the angular frequency of the wave, $\hat e_x$ is the unit vector pointing along the positive $x$-axis, $\Psi$ is a solution of the Schr\"odinger equation,
    \bea
    &&-\Psi''(z)+V(z)\Psi(z)=\frac{\omega^2}{c^2}\Psi(z),
    \label{s5-01}\\
    &&V(z):=\left\{\begin{array}{ccc}
    \displaystyle
    \frac{\omega^2[1-n(\omega,z)^2]}{c^2}&{\rm for}& |z|\leq L/2,\\ &&\\
    0 &{\rm for}& |z|> L/2,\end{array}\right.
    \label{s5-02}
    \eea
$c$ is the speed of light in vacuum, and $n(\omega,z)$ is the complex refractive index of the medium. As shown in Ref.~\cite{pra-2011b}, by a simple change of variables $z$ and $\omega$, namely
    \be
    z\to x:=\frac{z}{L}+\frac{1}{2},~~~~~\omega\to k:=\frac{L\omega}{c},
    \label{change}
    \ee
we can map (\ref{s5-01}) and (\ref{s5-02}) to (\ref{sch-eq}) and (\ref{v=}), respectively. This means that we can use our general results to compute the effects of inhomogeneity of the gain medium ($z$-dependence of the refractive index) provided that we express the perturbation parameter $\epsilon$, the coupling constants $\fz_1$ and $\fz_2$, and the function $f$ entering (\ref{5-1}) in terms of the physical parameters of the system. This requires the knowledge of the dispersion relation that determines the dependence of the complex refractive index on $\omega$ and $z$.

If we consider a gain medium that is obtained by doping a host medium of refraction index $n_0$ and modeled as a two-level atomic system with lower and upper level population densities $N_l$ and $N_u$, resonance frequency
$\omega_0$, and damping coefficient $\gamma$, then
    \be
    n^2(\omega,z)= n_0^2-
    \frac{\hat\omega_p(z)^2}{\hat\omega^2-1+i\hat\gamma\,\hat\omega},
    \label{epsilon}
    \ee
where $\hat\omega:=\omega/\omega_0$, $\hat\gamma:=\gamma/\omega_0$,
$\omega_p^2:=(N_l-N_u)e^2/(m_e\varepsilon_0)$, $e$ is electron's
charge, and $m_e$ is its mass. Furthermore, we have \cite{pra-2011a}
    \be
    \hat\omega_p(z)^2= 2\hat\gamma\kappa (z)\sqrt{n_0^2+\kappa (z)^2}
    \approx
    2\hat\gamma n_0\kappa (z)=-\frac{n_0\hat\gamma\, \lambda_0
    g (z)}{2\pi},~~~~~~~
    \kappa (z):=-\frac{\lambda_0 g (z)}{4\pi},
   \label{plasma2}
    \ee
where $\lambda_0:=2\pi c/\omega_0$ is the resonance wavelength,
$g (z)$ is the effective gain
coefficient (gain coefficient minus loss
coefficient) at the resonance frequency, and we have made use of the fact that for all known gain media, $|\kappa (z)|\ll n_0$. Substituting (\ref{plasma2}) in (\ref{epsilon}) gives
    \be
    n^2(\omega,z)= n_0^2+
    \frac{n_0\hat\gamma\,\lambda_0g(z)}{2\pi
    (\hat\omega^2-1+i\hat\gamma\,\hat\omega)},
    \label{n2=}
    \ee

Next, we assume that the pumping intensity decays exponentially inside the sample. This implies that if we pump it from the left-hand side, we have  \cite{pra-2011b},
    \be
    g(z)= (g_0+\alpha)\, e^{-\nu(\frac{z}{L}+\frac{1}{2})}-\alpha
    ~~~~{\rm for}~~~~|z|<\mbox{$\frac{L}{2}$},
    \label{g=1}
    \ee
where $g_0:=g(-\frac{L}{2})$, $\alpha$ is the attenuation coefficient at the resonance frequency that coincides with the largest allowed value of $g_0$, and $\nu$ is the decay constant that specifies the exponential decay of the intensity of the pumping beam inside the slab. If we pump the sample from both sides (double pumping), we instead find \cite{pra-2011b}
    \be
     g(z)=\left[\frac{g_0+\alpha}{\cosh(\frac{\nu}{2})}\right]
     \cosh(\mbox{\large$\frac{\nu z}{L}$})-\alpha
     ~~~~{\rm for}~~~~|z|<\mbox{$\frac{L}{2}$}.
    \label{g=2}
    \ee

Putting all these information together and making the change of variable (\ref{change}), we can reduce the problem of finding the OSSs of the above system to locating the spectral singularities of the potential (\ref{v=}) with $v_0$ and $v_1$ given by (\ref{5-1}) and the following choices for $\fz_1$, $\fz_2$, $\epsilon$, and $f(x)$, \cite{pra-2011b}.
    \bea
    \fz_1&:=&\left(\frac{2\pi L\hat\omega}{\lambda_0}\right)^2\left[
    1-n_0^2+\frac{\hat\gamma n_0\lambda_0 g_0\fz}{2\pi}\right],~~~~~~~~\fz_2:=2\pi\hat\omega^2\left(\frac{g_0}{\alpha}+1
    \right)\fz,
    \label{5-551}
    \\
    \epsilon&:=&\left\{\begin{array}{cc}
    \lambda_0^{-1}L^2\alpha\hat\gamma n_0\nu& \mbox{for single pumping},\\
    &\\
    \lambda_0^{-1}L^2\alpha\hat\gamma n_0\nu^2 & \mbox{for double pumping},\end{array}\right.
    \label{eps=}\\
    f(x)&:=&\left\{\begin{array}{cc}
    \displaystyle
    \frac{e^{-\nu x}-1}{\nu}& \mbox{for single pumping},\\ & \\
    \displaystyle
    \frac{\cosh[\nu(x-\frac{1}{2})]-\cosh(\frac{\nu}{2})}{\nu^2
    \cosh(\frac{\nu}{2})}& \mbox{for double pumping},\end{array}\right.
    \label{5-f=}
    \eea
where we have introduced:
    \be
    \fz:=\frac{1}{1-\hat\omega^2-i\hat\gamma\hat\omega}.
    \label{5-812}
    \ee
We also note that according to (\ref{fn=}) and (\ref{5-551}),
    \be
    \fn=\sqrt{n_0^2-\frac{\hat\gamma n_0\lambda_0 g_0\fz}{2\pi}}=
    \sqrt{n_0^2-\frac{\hat\gamma n_0\lambda_0 g_0\lambda^2}{2\pi(\lambda^2-i\hat\gamma\lambda_0\lambda-\lambda_0^2)}},
    \ee
where $\lambda$ stands for the wavelength of the wave, i.e.,
    \be
    \lambda:=\frac{2\pi c}{\omega}=\frac{2\pi L}{k}=\frac{\lambda_0}{\hat\omega},
    \label{wl=}
    \ee

Because the physical parameters of practical interest are the wavelength $\lambda$ and the gain coefficient $g_0$, we fix all the other physical quantities and study the effect of inhomogeneity of the medium on the $\lambda$ and $g_0$ values associated with spectral singularities. These we respectively denote by $\lambda_\star$ and $g_{\star}$ and expand in power series in the perturbation parameter $\epsilon$,
    \be
    \lambda_\star=\lambda_{(0)}+\sum_{m=1}^\infty \lambda_{m}\epsilon^m,~~~~~~~~~~~
    g_{\star}=g_{(0)}+\sum_{m=1}^\infty g_{m}\epsilon^m.
    \label{w5-04}
    \ee
Here $\lambda_{(0)}$ and $g_{(0)}$ are respectively the values of $\lambda$ and $g_0$ associated with the spectral singularity of the unperturbed potential that appears for $\fn=\fn_0$ and $k=k_0$. As shown in Ref.~\cite{pra-2011a}, these are labeled by a mode number $m$ and can be computed by substituting (\ref{n2=}) and $\hat\omega=\lambda_0/\lambda$ in (\ref{ss-zero}) and finding the real values of $\lambda$ and $g_0$ that satisfy (\ref{ss-zero}). In view of (\ref{5-19}) and (\ref{wl=}), it is easy to see that
    \be
    k_0=\frac{2\pi L}{\lambda_{(0)}},~~~~~~~
    k_1=-\frac{2\pi L\lambda_1 }{\lambda_{(0)}^2}=-\frac{k_0\lambda_1}{\lambda_{(0)}}.
    \label{kk=}
    \ee
Next, we view $\fn$ as a function of $\lambda$ and $g_0$, and identify it with its Taylor series about $(\lambda_{(0)},g_{(0)})$. In light of (\ref{w5-04}), this gives for the value of $\fn$ at $(\lambda_\star,g_\star)$ the second equation in (\ref{5-19}), i.e., $\fn_\star=\sum_{m=0}^\infty\fn_m\epsilon^m$, and allows for identifying the coefficients $\fn_m$ in terms of those of the power series (\ref{w5-04}) for $\lambda_\star$ and $g_\star$. In particular, $\fn_0$ is the value of $\fn$ at $(\lambda_{(0)},g_{(0)})$, i.e.,
    \bea
    &&\fn_0:=\sqrt{1-\frac{\fz_{(1)}}{k_0^2}},
    \label{5-329}\\
    &&\fz_{(1)}:=\fz_1 \Big|_{\lambda=\lambda_{(0)},g_0=g_{(0)}}=
    -\left(\frac{2\pi L}{\lambda_{(0)}}\right)^2
    \left[n_0^2-1-\frac{\hat\gamma n_0\lambda_0\lambda_{(0)}^2g_{(0)}}{
    2\pi(\lambda_{(0)}^2-i\hat\gamma\lambda_0\lambda_{(0)}-\lambda_0^2)}\right],
    \label{5-330}
    \eea
and $\fn_1$ is given by
    \bea
    \fn_1&:=&\fn_{1,0}\lambda_1+\fn_{0,1}g_1,
    \label{5-331}\\
    \fn_{1,0}&:=&\left.\frac{\partial \fn
    }{\partial \lambda}\right|_{
    \lambda=\lambda_{(0)},\,g_0=g_{(0)}}=
    \frac{\hat\gamma n_0\lambda_0^2\lambda_{(0)}
    (2\lambda_0+i\hat\gamma\lambda_{(0)})g_{(0)}}{4\pi\fn_0(
    \lambda_{(0)}^2-i\hat\gamma\lambda_0\lambda_{(0)}-\lambda_0^2)^2},
    \label{5-678}
    \\
     \fn_{0,1}&:=&\left.\frac{\partial \fn
    }{\partial {g_0}}\right|_{
    \lambda=\lambda_{(0)},\,g_0=g_{(0)}}=
    \frac{-\hat\gamma n_0\lambda_0\lambda_{(0)}^2}{4\pi\fn_0(
    \lambda_{(0)}^2-i\hat\gamma\lambda_0\lambda_{(0)}-\lambda_0^2)}.
    \label{5-679}
    \eea

Now, we recall that the spectral singularities are obtained from the equation
    \be
    \Gamma_{1-}=\sum_{\ell=1}^\infty F_\ell\,\fz_2^\ell\,\epsilon^\ell=0.
    \label{5-343}
    \ee
In order to use perturbation theory to solve this equation for $\lambda$ and $g_0$, we need to express $F_\ell$ and $\fz_2$ as functions of $\lambda$, $g_0$, and $\epsilon$, substitute their Taylor series expansion about $(\lambda_{(0)},g_{(0)})$ in $\sum_{\ell=1}^\infty F_\ell\fz_2^\ell\epsilon^\ell$, and then set the coefficients of the resulting powers series to zero. As we will see below, for most realistic situations the first-order perturbation theory gives highly reliable results. Therefore, we outline the details of the calculation of the first-order corrections to $\lambda_{(0)}$ and $g_{(0)}$, namely $\lambda_1$ and $g_1$.

Determination of $\lambda_1$ and $g_1$ requires the following expansions of
$F_0$, $F_1$, and $\fz_2$.
    \be
    F_0=F_{(0)}+(X\lambda_1+Yg_1)\epsilon+\cO(\epsilon^2),~~~~
    F_1=F_{(1)}+\cO(\epsilon^1),~~~~
    \fz_2=\fz_{(2)}+\cO(\epsilon^1),
    \label{5-344}
    \ee
where $\cO(\epsilon^\ell)$ stands for the terms of order $\epsilon^\ell$ and higher, and
    \bea
    F_{(0)}&:=&F_0(\fn_0,k_0)=F_{000},~~~~~~~
    X:=\fn_{10}F_{010}-\frac{2\pi L F_{001}}{\lambda_{(0)}^2},
    \label{5-345}\\
    Y&:=&\fn_{01}F_{010},~~~~~~~
    F_{(1)}:=F_1(\fn_0,k_0)=F_{100},
    \label{5-345b}\\
    \fz_{(2)}&:=&\fz_{2}\Big|_{
    \lambda=\lambda_{(0)},\,g_0=g_{(0)}}=
    \frac{2\pi\lambda_0^2(g_{(0)}+\alpha)}{\alpha
    (\lambda_{(0)}^2-i\hat\gamma\lambda_0\lambda_{(0)}-\lambda_0^2)},
    \label{5-346}
    \eea
In the derivation of these formulas we have made use (\ref{5-19}), (\ref{5-21}), (\ref{5-551}), (\ref{5-812}), (\ref{kk=}), and (\ref{5-331}).

Next, we observe that because for $(\fn,k)=(\fn_0,k_0)$ we have a spectral singularity, $F_{(0)}=0$. In view of this relation,  (\ref{5-4}), and (\ref{5-21}),  we can calculate
    \be
    F_{010}=\frac{\fz_{(1)}+i2k_0}{\fn_0}=
    \frac{1}{\fn_0}\left[\fz_{(1)}+\frac{4\pi i L}{\lambda_{(0)}} \right],~~~~~~~
    F_{001}=-k_0(\fn_0^2-1)=
    \frac{\fz_{(1)}}{k_0}=\frac{\lambda_{(0)}\fz_{(1)}}{2\pi L}.
    \label{5-357}
    \ee
These together with (\ref{5-345}) and (\ref{5-346}) give
    \be
    X=\frac{\fn_{10}}{\fn_0}\left[\fz_{(1)}+\frac{4\pi i L}{\lambda_{(0)}}\right] -\frac{\fz_{(1)}}{\lambda_{(0)}},~~~~~~
    Y=\frac{\fn_{01}}{\fn_0}\left[\fz_{(1)}+\frac{4\pi i L}{\lambda_{(0)}}\right].
    \label{5-XY=}
    \ee
Furthermore, in light of $F_{(0)}=0$ and (\ref{5-343}) -- (\ref{5-346}), we can write the equation determining $\lambda_1$ and $g_1$ as
    \be
    X\lambda_1+Yg_1=-F_{100}\,\fz_{(2)}.
    \label{5-347}
    \ee
This is a complex linear equation involving two real unknowns. Therefore, we can easily solve it to obtain:
    \be
    \lambda_1=-\frac{\IM(F_{100}\,\fz_{(2)}Y^*)}{\IM(XY^*)},~~~~~~~
    g_1=\frac{\IM(F_{100}\,\fz_{(2)}X^*)}{\IM(XY^*)},
     \label{5-348}
    \ee
where ``$\IM(\cdot)$'' stands for the imaginary part of its argument.

Next, we examine $F_{100}$ that encodes all the information about the inhomogeneity of the gain medium. According to (\ref{5-5b}),
    \be
    F_{100}=F_1(\fn_0,k_0)=\frac{1}{\fn_0\, k_0}\int_0^1 dx\, \xi(\fn_0,k_0,x)f(x),
    \label{5-632}
    \ee
where $\xi$ is given by (\ref{xi=}). With the help of
(\ref{ss-zero}) we have been able to simplify the expression for $\xi(\fn_0,k_0,x)$ and find
    \be
    \xi(\fn_0,k_0,x)=\left(1-\frac{1}{\fn_0^2}\right) \cos^2\left[\fn_0k_0(x-\mbox{$\frac{1}{2}$})\right].
    \label{5-633}
    \ee
Substituting this equation and (\ref{5-f=}) in (\ref{5-632}), evaluating the resulting integrals, and using (\ref{ss-zero}) to simplify the outcome we obtain
    \be
    F_{100}=\frac{(1-\frac{1}{\mf^2})\Big[(1-e^{-\nu})[4i+k_0(1+\mf^2)]-k_0\nu+
    2i\nu^3-4k_0^3\;\mf^3(\mf^2-1)(\nu+e^{-\nu}-1)     \Big]}{2k_0^3\;\mf^4\;\nu^2(4k_0^2\;\mf^2+\nu^2)},
    \label{F100=1}
    \ee
for a singly-pumped sample, and
    \be
    F_{100}=\frac{(\mf-1)(\mf+1)^2[-2+\mf^2(\nu-2)-\mf\nu]}{2k_0^2\;
    \mf^7\;\nu(4k_0^2\;\mf^2+\nu^2)},
    \label{F100=2}
    \ee
for a doubly-pumped sample.

Now, we can use (\ref{eps=}), (\ref{5-330}), (\ref{5-XY=}), (\ref{5-348}), (\ref{F100=1}), and (\ref{F100=2}) to calculate the first-order corrections to the wavelength and gain coefficient of the OSSs, i.e., $\lambda_1\epsilon$ and $g_1\epsilon$.

As a concrete example, consider a semiconductor gain medium with the following specifications that is also studied in Refs.~\cite{pra-2011a,pra-2011b}.
    \be
    n_0=3.4,~~~~L=300~\mu{\rm m},~~~~
    \lambda_0=1500~{\rm nm},~~~~
    \hat\gamma=0.02,~~~~
    \alpha=200~{\rm cm}^{-1},~~~~0\leq\nu\leq 0.5.
    \label{specific=}
    \ee
For this sample $\epsilon\,\approx 0.8\, \nu\leq 0.4$ for single pumping and $\epsilon\approx 0.8\,\nu^2\leq 0.2$ for double pumping. In particular for physically realistic situations where $\nu\precsim 0.1$, we have $\epsilon^2\precsim 6.4\times 10^{-3}$ and $\epsilon^2\precsim 4.0\times 10^{-4}$ for single and double pumping, respectively. This shows that the first-order perturbation theory produces a highly reliable description of the OSSs for this system.

In the remainder of this section, we report the results of the first-order perturbative calculation of $g_\star$ and $\lambda_\star$ for the first five OSSs that appear as we increase the intensity of the pumping beam starting from zero. These have a wavelength that is closest to the resonance wavelength of the sample, $\lambda_0=1500~{\rm nm}$. We label them using the mode number $m$ that is introduced in \cite{pra-2011b} and takes values between 1358 and 1362, with 1360 corresponding to the OSS with closest wavelength to $\lambda_0$. Table~\ref{table1} gives numerical values of $\lambda_\star$ and $g_\star$ for these OSSs and five different values of the decay constant $\nu$. As we increase $\nu$, $\lambda_\star$ remains essentially unchanged while $g_\star$ increases. This is particularly pronounced for a singly-pumped sample. It confirms the semiclassical results obtained in \cite{pra-2011b}.
    \begin{table}
    \begin{center}
    \begin{tabular}{|c|c||c|c||c|c|}
    \hline
    \multicolumn{2}{|c||}{}&\multicolumn{2}{|c||}{Single Pumping} &\multicolumn{2}{|c|}{Double Pumping}\\
    \hline \hline
    m&$\nu$&$\lambda\;(\text{nm})$&$\mg\;(\text{cm}^{-1})$&
    $\lambda\;(\text{nm})$&$\mg\;(\text{cm}^{-1})$\\
    \hline
    \multirow{5}{*}{1362}
    &0.0&1497.561770810&41.53101&1497.561770810&41.53101\\
    &0.1&1497.561770785&43.45261&1497.561770784&41.56407\\
    &0.2&1497.561770716&45.25128&1497.561770707&41.66286\\
    &0.3&1497.561770609&46.93581&1497.561770579 &41.82620\\
    &0.5&1497.561770304&49.99447&1497.561770180&42.33818\\ \hline
    \multirow{5}{*}{1361}
    &0.0&1498.389018373&40.91032&1498.389018373&40.91032\\
    &0.1&1498.389018341&42.83283&1498.389018339&40.94324\\
    &0.2&1498.389018251&44.63206&1498.389018239&41.04159\\
    &0.3&1498.389018115&46.31686&1498.389018073&41.20423\\
    &0.5&1498.389017715&49.37529&1498.389017554&41.71399\\ \hline
    \multirow{5}{*}{1360}
    &0.0&1499.999983312&40.40905&1499.999983312&40.40905\\
    &0.1&1499.999983275&42.33379&1499.999983220&40.44217\\
    &0.2&1499.999983205&44.13541&1499.999983115&40.54115\\
    &0.3&1499.999983098&45.82273&1499.999983003&40.70480\\
    &0.5&1499.999982791&48.88649&1499.999982512&41.21777\\ \hline
    \multirow{5}{*}{1359}
    &0.0&1501.475689102&40.79650&1501.475689102&40.79650\\
    &0.1&1501.475689077&42.72315&1501.475689075&40.82968\\
    &0.2&1501.475689007&44.52660&1501.475688997&40.92881\\
    &0.3&1501.475688899&46.21566&1501.475688689&41.09272\\
    &0.5&1501.475688590&49.28266&1501.475688464&41.60649\\ \hline
    \multirow{5}{*}{1358}
    &0.0&1502.670951310&41.63220&1502.670951310&41.63220\\
    &0.1&1502.670951286&43.56043&1502.670951282&41.65321\\
    &0.2&1502.670951220&45.36542&1502.670951211&41.76466\\
    &0.3&1502.670951118&47.05600&1502.670951089&41.92890\\
    &0.5&1502.670950826&50.12593&1502.670950707&42.44373\\ \hline
    \end{tabular}
    \vspace{6pt}
    \caption{The wavelength $\lambda_\star$ and gain coefficient $g_\star$ for the five OSSs that are generated by pumping the semiconductor gain medium (\ref{specific=}). $m$ is the mode number labeling these OSSs \cite{pra-2011b}. $\nu$ is the decay constant for the intensity of the pumping beam(s) inside the sample.}
    \label{table1}
    \end{center}
    \end{table}

\begin{figure}
    \begin{center}
    \includegraphics[scale=.6]{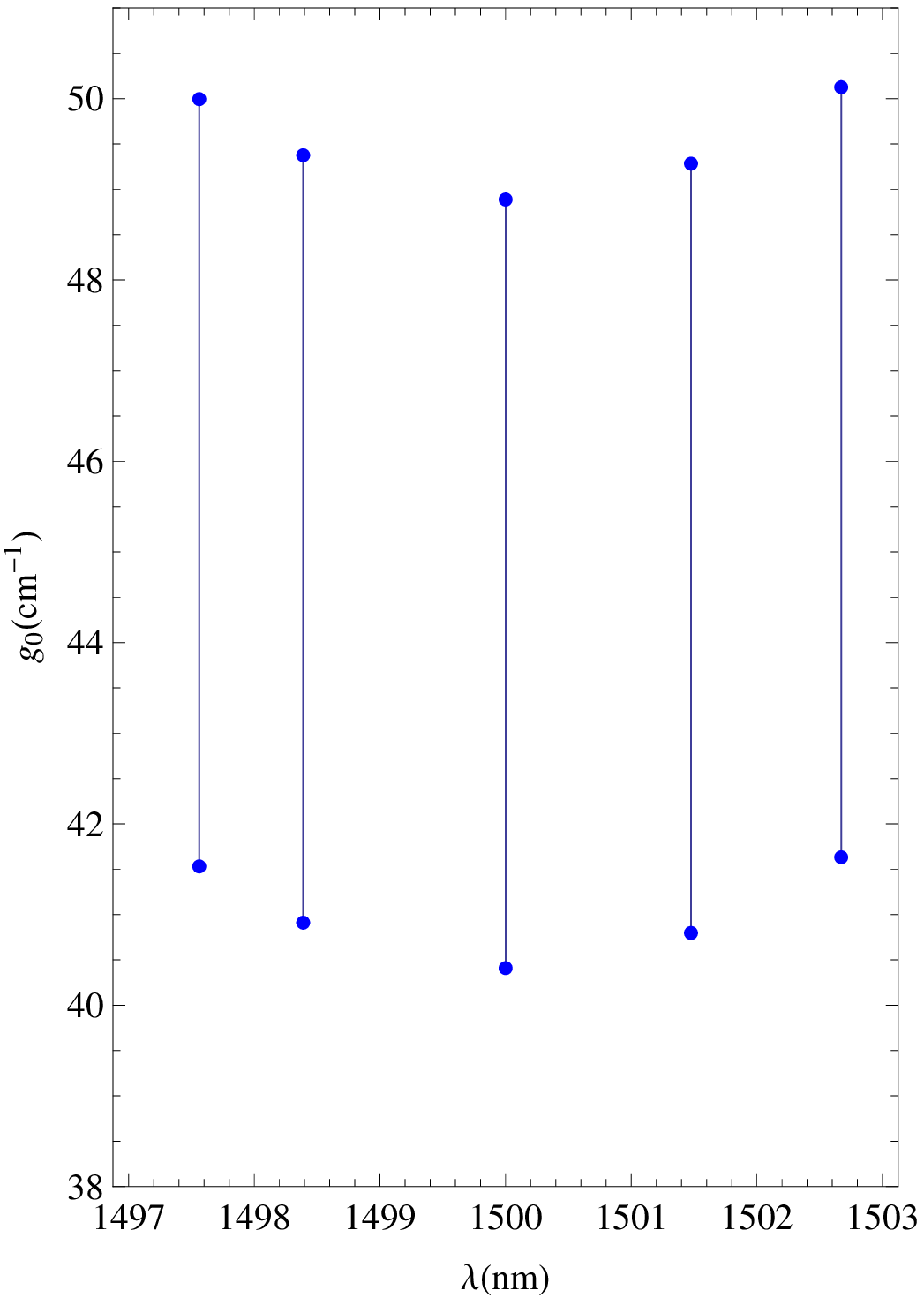}%
    \hspace{1cm}%
    \includegraphics[scale=.6]{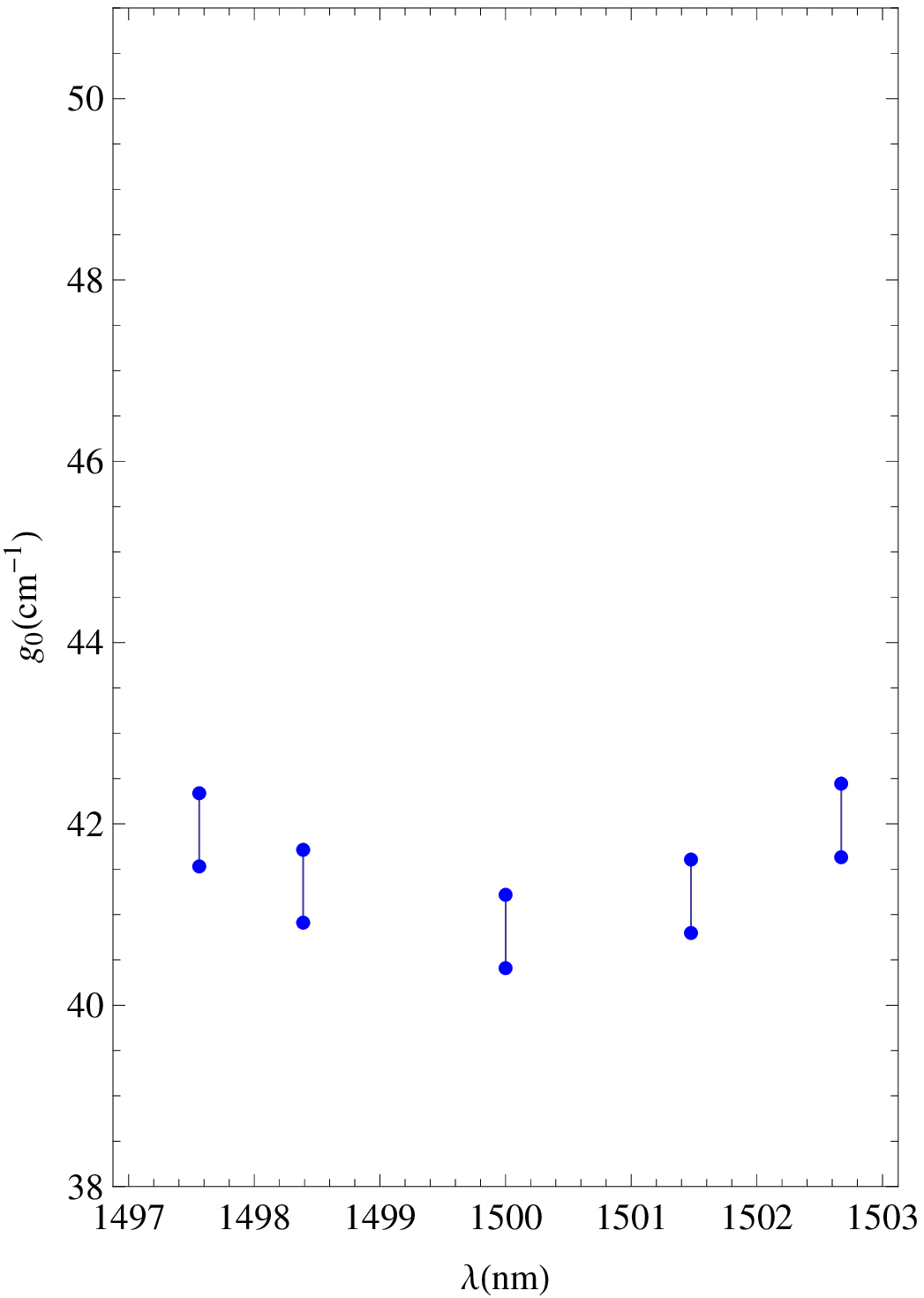}
    \caption{(Color online) Location of the optical spectral singularities considered in Table~\ref{table1} for $0\leq\nu\leq 0.5$. The left- and right-hand figures
    respectively show the behavior of OSS for the singly and doubly pumped samples. The displayed dots that mark the boundaries of each of the curves of the OSSs correspond to $\nu=0$ and $\nu=0.5$. As we increase $\nu$ the location of OSSs in the $\lambda$-$g_0$ plane moves upwards on the given curves.}
    \label{fig2}
    \end{center}
    \end{figure}

Figure~\ref{fig2} shows the curves traced by $(\lambda_\star,g_\star)$ as we change $\nu$ for each of the OSSs considered in Table~\ref{table1}. The fact that these curved are essentially vertical line segments shows that $\lambda_\star$ does not depend on $\nu$, while the opposite is true for $g_\star$, especially for the singly-pumped sample. The displayed dots that mark the lower and upper boundaries of each line segment respectively correspond to $\nu=0$ and $\nu=0.5$. As we increase $\nu$, the location of each OSS in the $\lambda$-$g_0$ plane moves upwards along the corresponding line segment.

\section{Concluding Remarks}

In this article we have outlined a general method for carrying out a perturbative calculation of the transfer matrix for a general complex potential that vanishes outside a closed interval. This allows for a systematic characterization of the spectral singularities of this class of potentials. This turns out to be particularly suitable for the study of optical spectral singularities and their time-reversed analogs that respectively correspond to lasing at threshold gain and antilasing.

As an application of our general results, we examined the problem of computing the transfer matrix for a potential consisting of finitely many Dirac delta-functions that are centered at arbitrary points and have arbitrary complex coupling constants. We showed that perturbation theory gives an exact expression for the transfer matrix for this system.

Next, we considered an arbitrary complex perturbation of a constant (complex) barrier potential and applied our method to compute the effect of the perturbation on the spectral singularities. A physical realization of this model is in the description of threshold lasing associated with an infinite planar slab of gain material. For this system the gain coefficient that is proportional to the intensity of the pumping beam decays exponentially as the beam penetrates the medium. This in turn makes the gain medium inhomogeneous. Our method allows for an essentially analytic calculation of the effect of this inhomogeneity on the location of optical spectral singularities. Our results confirm those obtained using the method of Ref.~\cite{pra-2011b} that is based on the semiclassical approximation. Compared with this method ours
has the advantage of being applicable in every spectral range. In particular, we can use it in the spectral ranges comparable with the length scale of the system. This is, for example, the case in the recent study of unidirectional invisibility \cite{unidir}. Our method allows for a thorough analysis of this and much more general optical systems displaying threshold lasing, antilasing, and unidirectional invisibility.

\vspace{.3cm}
\noindent {\em Acknowledgments:} This work has been
supported by  the Scientific and Technological Research Council of
Turkey (T\"UB\.{I}TAK) in the framework of the project no: 110T611 and the
Turkish Academy of Sciences (T\"UBA). We wish to thank Aref Mostafazadeh for his help in preparing Figure~\ref{fig1} and Ali Serpeng\"{u}zel  for his careful reading of the first draft of this paper and making many invaluable remarks.

\ed
\begin{thebibliography}{99}
\bibitem{endnote-x1} For sufficiently large $\epsilon$, the presence of $\epsilon\, v_1$ may also lead to the creation of new spectral singularities.
\end{thebibliography}
